\documentclass[]{spie}  

\usepackage{amsmath,amsfonts,amssymb}
\usepackage{graphicx}
\usepackage[colorlinks=true, allcolors=blue]{hyperref}
\usepackage{subcaption}

\title{RF Sensing for Continuous Monitoring of Human Activities for Home Consumer Applications}

\author[a]{Moeness G. Amin}
\author[b]{Arun Ravisankar}
\author[a]{Ronny G. Guendel}
\affil[a]{Center for Advanced Communication, Villanova University, Villanova, PA 19085, USA}
\affil[b]{Comcast Labs, Comcast, Philadelphia, PA 19103, USA}

\begin{document} 
\maketitle

\begin{abstract}
		Radar for indoor monitoring is an emerging area of research and development, covering and supporting different health and wellbeing applications of smart homes, assisted living, and medical diagnosis. We report on a successful RF sensing system for home monitoring applications. The system recognizes Activities of Daily Living(ADL) and detects unique motion characteristics, using data processing and training algorithms. We also examine the challenges of continuously monitoring various human activities which can be categorized into translation motions (active mode) and in-place motions (resting mode). We use the range-map, offered by a range-Doppler radar, to obtain the transition time between these two categories, characterized by changing and constant range  values, respectively. This is achieved using the Radon transform that identifies straight lines of different slopes in the range-map image. Over the in-place motion time intervals, where activities have insignificant or negligible range swath,  power threshold of the radar return micro-Doppler signatures,which is employed to define the time-spans of individual activities with insignificant or negligible range swath. Finding both the transition times and the time-spans of the different motions  leads to improved classifications, as it avoids decisions rendered over time windows covering mixed activities.
\end{abstract}

	\keywords{Radar, indoor monitoring, smart homes, motion classification}

\section{INTRODUCTION}
\label{sec:intro}  
With our aging population, and the desire for independence, the need to monitor Activities of Daily Living (ADL) is a growing problem in society.
\cite{1_aminSP, 2_amin2017radar} As life expectancy increases, so will the demand for elders to age in place as opposed to relocating to assisted living environments.  Aging in place describes empowering seniors with the ability to live in one’s own home, safely and independently.  Since decline in health is inevitable, aging-in-place requires increasing amounts of support to maintain an independent existence.  The combination of sensing technologies and data analytics are powerful tools in providing support to those caring for seniors who are living independently.  Sensors provide insights that characterize the potential for living a quantified life.  To compress morbidity, healthcare professionals will benefit from the insights provided from continuous monitoring of both classified medical data, such as vitals, and lifestyle monitoring of activities of daily living.  Activities of daily living are the routine matters of our day, most notably including eating, sleeping, movement, hygiene and toileting.  Fitness and cognitive activities are also strong indicators of state of health.

\begin{figure}[hbtp] 
		\centering
		\includegraphics[width=0.9\linewidth]{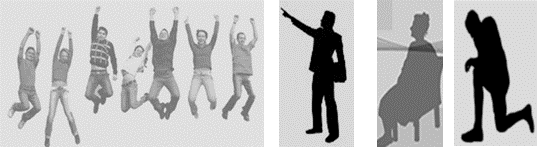}
				\vspace{1em}
		\caption{In-place motions, jumping (left) arm gesture and sitting (middle), kneeling (right)}
		\label{fig:Into2IP-Motion}
    \end{figure}

\begin{figure}[hbtp] 
		\centering
		\includegraphics[width=0.9\linewidth]{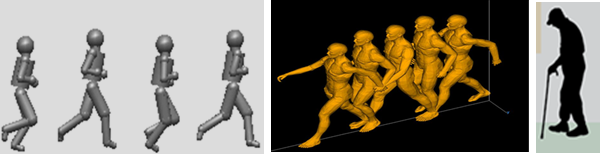}
		\vspace{1em}
		\caption{Translation motions, running (left), walking (middle), slow walking (right)}
		\label{fig:Into1TranslMotion}
    \end{figure}

\begin{figure}[hbtp] 
		\centering
		\includegraphics[width=0.9\linewidth]{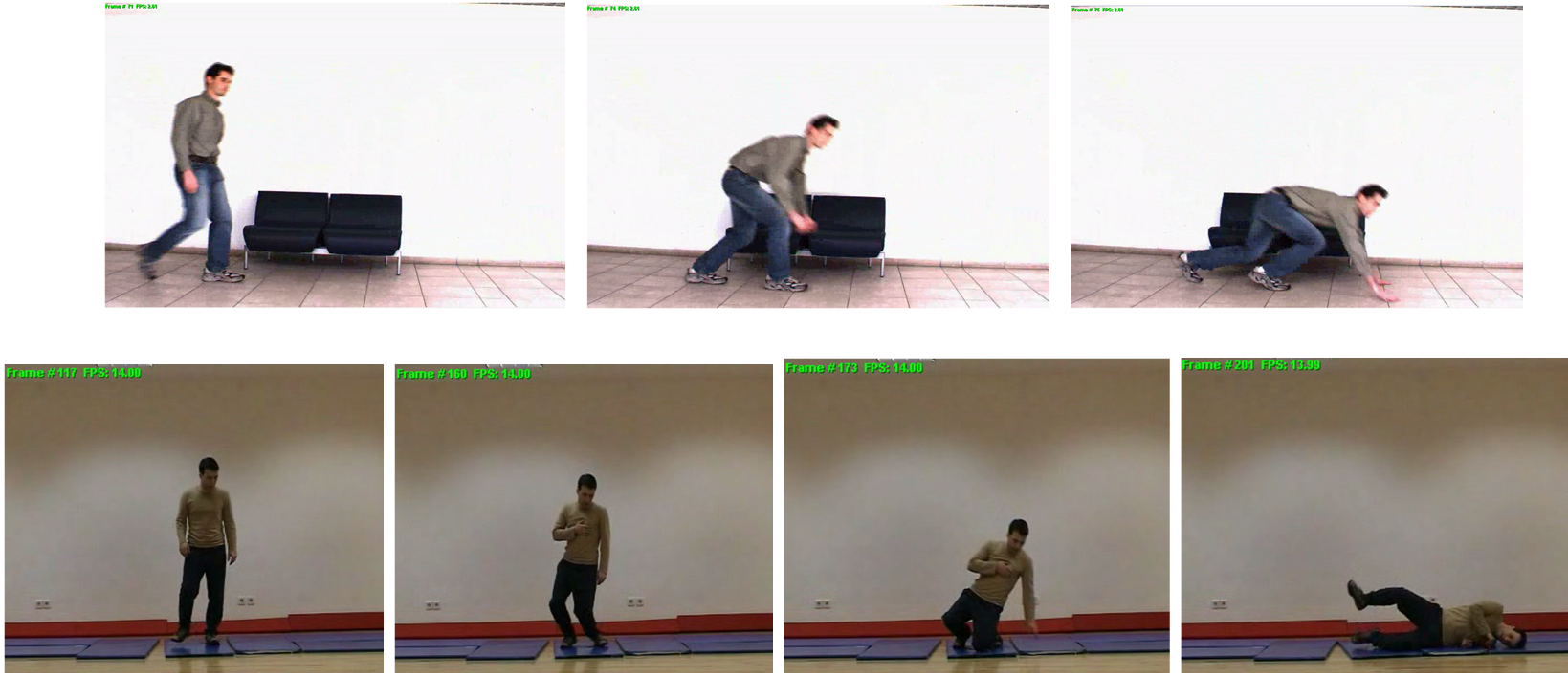}
				\vspace{1em}
		\caption{Fall can be either considered a translation motion (top) or an in-place (bottom) motion}
		\label{fig:Into3Falling}
	\end{figure}

Apart from the health and wellness applications, RF-based sensors promise non-intrusive, safe and efficient home automation/security applications.\cite{3_7842648, 4_7952908,5_7944373,6_seifert2019RadarConf,7_6875709,8_4801689,9_7348882,10_5404249, 11_bijan2009} Falls are considered as an abnormal activity that should be accurately detected and classified with high sensitivity and specificity.\cite{12_Hong2011, 13_Su2015, 14_Wu2015} Gesture recognition/control has great potential in enhancing Man Machine Interface (MMI). The ability of radar-based sensors to precisely detect motion as well as describe unique characteristics of a motion activity provides high fidelity sensor data to the home security/automation system. The penetration of RF signals through walls gives added advantage.  Radar offers fine controls for sensing configurations (selectively choosing what to track; ignoring targets that are of no interest). It works in all lighting and  weather conditions.\cite{15_amin2016through, 16_taylor1994introduction, 17_skolnik2002introduction, 18_chenbook} 
Human daily activities can be further categorized into translation and in-place motions. Whereas the former mainly describes crawling and gait articulations, via normal or abnormal walking, or running actions, the latter is primarily associated with motions that do not exhibit considerable changes in range.  In-place motions include sitting, standing, kneeling, and bending: each is performed without any stride. The accumulated range under each  in-place  is typically less than 1 meter.  Fall can be considered a translation or in-place motion, depending on its type.  Whereas “progressive” fall exhibits a large range span, “drop” like fall is confined to a relatively small range extent. Examples of translation motions are given in Fig.~\ref{fig:Into1TranslMotion} which depicts two extremes of fast and slow walking speeds. Fig.~\ref{fig:Into2IP-Motion} shows cases of in-place motions of jumping, arm gestures, and sitting.  Fig.~\ref{fig:Into3Falling} shows two types of falls pertaining to two different fall categories.
	
The contribution of this paper is two-fold. We report the working of a successful RF sensor for home applications, which is jointly developed by Villanova University with support from Comcast Labs,  the advanced technology arm of the company where ideas, technologies and user experiences are tested, evaluated and refined. It is important to point out that from a Comcast Labs point of view, this is a technological test, not a product, and if any productization of this test would only occur, it would only be with the full, opt-in consent of the individual being monitored. That said, it is also our view that a radar-based approach for in-home monitoring plausibly provides peace of mind for both the person being monitored, and the caregiver, for the following reasons: 1) The intended customer can “age in place” at home for a longer period of time, without the use of cameras to monitor activity and 2) Caregivers (often adult children) receive peace-of-mind, also without having to rely on cameras to “look in on” their charge(s). We also introduce a simple method based on the Radon transform, applied to the range-map, for separating translation from in-place motions.  The range-map depicts the range vs slow-time of a moving target. Accordingly, horizontal lines correspond to in-place motions, whereas translation motions are manifested by lines with non-zero slopes. We also remark that, in the range-map, fast translation motions are associated with steep lines. For example, running has higher slopes than walking, and walking has a higher slope than crawling.  Further, elderly gait typically leave behind smaller signature slopes compared to those of young adults. Translation motion stemming from using a wheelchair is comparable with an average walking speed, producing a similar range slope. Positive slopes describe range translation away from radar, and negative slopes are a result of walking towards the radar. It is also noted that the range-map does not exhibit abrupt, sudden step change behavior, like that of square waveforms. In essence, it consists of piece-wise straight lines with finite slopes.

\section{Activity classification System}
\label{sec:2_ActClassSystUndertaking}  
\begin{figure}[hbtp] 
	\centering
	\includegraphics[width=0.8\linewidth]{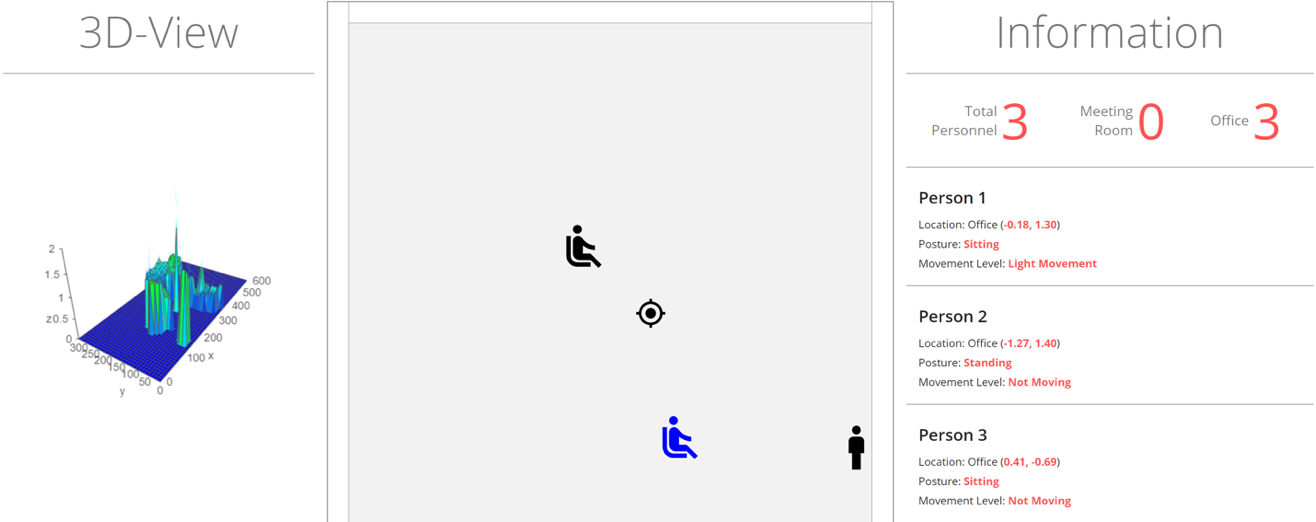}
			\vspace{1em}
	\caption{Radar application showing presence and postures.}
	\label{fig:SecondSec_RadarAppl}
\end{figure}

With more than a decade spent by many researchers worldwide to develop algorithms and device motion and fall classification techniques, it has become vital - now more so than ever before - to transition these efforts into a working system. In 2018, Comcast Labs and Villanova University embarked on Phase-I of a project which aimed at identifying key and effective algorithms that could be applied to radar for indoor monitoring, without the use of cameras. The results were presented at the 2018-SPIE.\cite{19_erol2018RadarFallcomcast}  The experiments discussed in that paper included a variety of motions.  The data was collected using a 24 GHz radar that was used to test a data-driven feature-learning algorithm, namely, two-dimensional principal component analysis (2D-PCA),\cite{20_8103916, 21_seifert2018EUSIPCO, 22_6252625, 23_4359192, 24_Jokanovic, 25_Ye2004GED10140521014092} followed by a nearest neighbor classifier. In essence, a complete motion classification application was built using both hardware and software components. Fig.~\ref{fig:SecondSec_RadarAppl} shows a potential application of a radar-based sensor installed on the wall and covering an area of the home.  In this case, the radar monitors the human inside the home and provides motion location and posture, and immediately displays this information on a wide screen. The system is also able to tell if the person exhibits no movement/light movement/heavy movement. This information could prove important for health providers and is useful in assessing deviations from average daily routines.

\begin{figure}
		\centering
		\begin{minipage}[b]{0.49\textwidth}
			\centering
			\includegraphics[width=\textwidth]{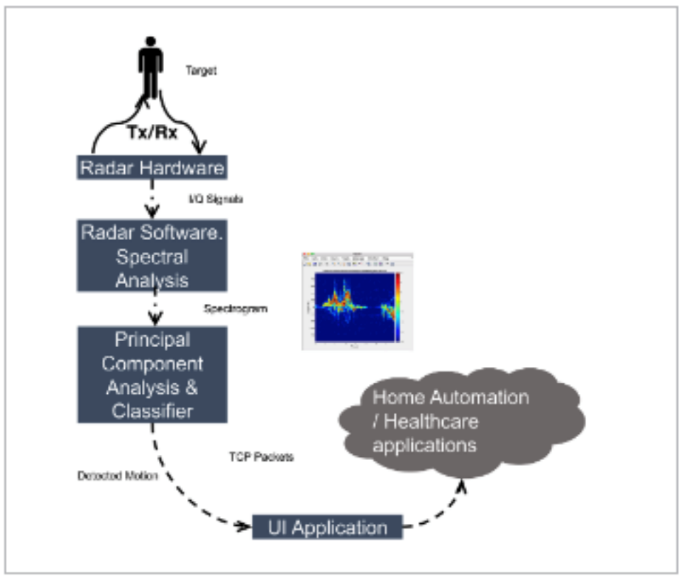}
			\vspace{0em}
			\caption{Steps involved in the activity classification project.}
			\label{fig:secondSecFlowGraph}
\end{minipage}
\begin{minipage}[b]{0.49\textwidth}
			\centering
			\includegraphics[width=\textwidth]{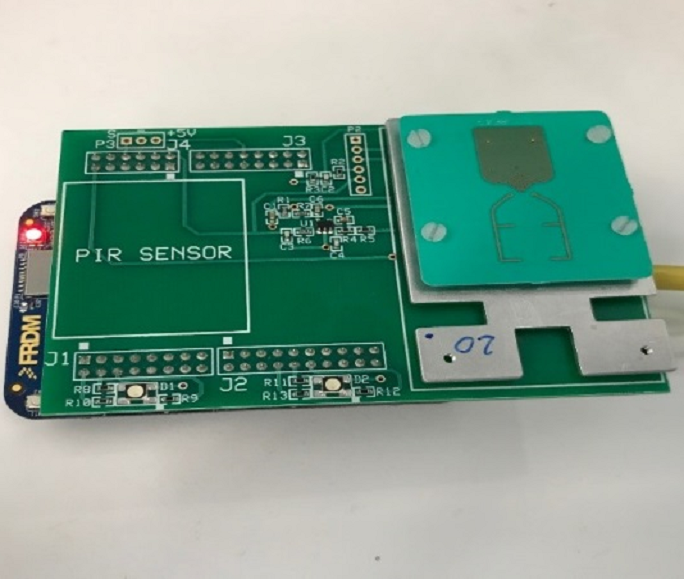}
			\vspace{0em}
			\caption{Radar used for Activity classification application.}
			\label{fig:radarSensor}
		\end{minipage}
\end{figure}

Fig.~\ref{fig:radarSensor} shows the image of the radar chip that was used in the project, which is now advancing to Phase-II. The radar antenna is connected to a micro-controller that does basic processing and gives out range and Doppler data. This information is then processed by 2D-PCA for feature extraction. The classifier would then identify the activity that is taking place in the radar arena. Activities include walking, sitting, standing, bending, gesture, and falling. 
	
\begin{figure}[hbtp] 
		\centering
		\includegraphics[width=0.99\linewidth]{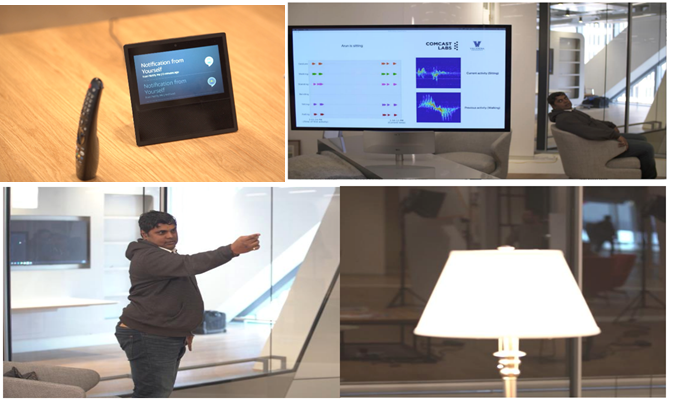}
		\vspace{1em}
		\caption{Fall detection notification sent to Alexa (Top Left), Application Detects Sitting (Top Right). Gesture control tuning on smart light (Bottom)}
		\label{fig:SecondSec_Demo}
	\end{figure}

Fig.~\ref{fig:secondSecFlowGraph} shows the steps involved in the project. In this system, the algorithm would detect motion and send a message to an application that would display as a dashboard. The spectrograms are generated from the radar returns. Feature extraction is followed by  kNN classifiers to determine the activity. Activity detected by the classifier is then sent to a python application which would be used to visualize the data. We also had gesture control to turn on/off home appliances. A notification would be sent to smart assistants, like Alexa, when the radar detects a fall. Fig.~\ref{fig:SecondSec_Demo} shows this application, and depicts current and past activities, as well as gesture control of smart home appliances.
	
	\section{Radon Transform for Range-Map Processing}

The Radon transform is considered an effective tool in detecting dominant color schemes in images, especially in medical image processing. The Radon transform is an integral transform method, while its inverse is typically used for reconstructing images from medical CT scans.\cite{29_wininger2013basis} In the underlying problem, the goal is to apply the Radon transform to the range-map of the radar signal returns to detect pertinent line structures. 
In employing the Radon transform for human motion recognition problems, we recognize that horizontal lines in the range-map correspond to in-place motions with no noticeable range extent, whereas lines with non-zero inclinations represent continuous changes in range gates stemming from motion translations, such as constant speed walking, crawling, or running. It is noted that acceleration or deceleration give rise to curvy signatures in the range-map where Hough transform can be applied.\cite{30_4599159}  In-place motions  include (a) sitting, (b) kneeling down from a standing position, (c) standing up from a kneeling or sitting position, and (d) bending while picking an object from the ground. 
Successful separations between translation and in-place classes of motions helps in identifying the "actions" vs "resting" modes at home.  Information in the range-map is sufficient to label the human activity as one of the two aforementioned classes.  In this respect, any follow-on classification of human motions using spectrograms for microDoppler signature estimation can proceed without confusing one class with another. For example, classifiers applied to time intervals of in-place motions will not have walking as a possible outcome.  Under the proposed framework, we process the range-map and microDoppler representations sequentially with the former preceding.

	\subsection{Application of the Radon Transform}
	\label{subsec:radonTrans}
\begin{figure}[hbtp]
		\centering
		\begin{subfigure}[b]{0.49\linewidth}
			\includegraphics[width=\linewidth]{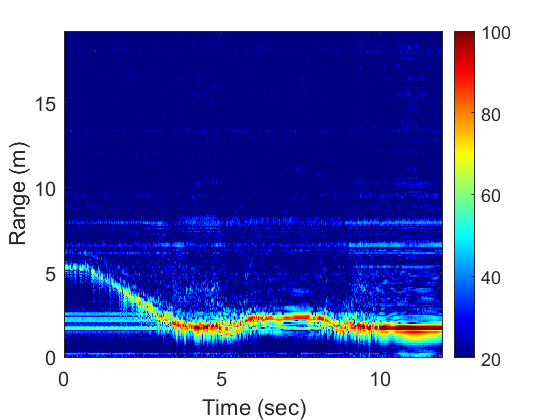}
			\caption{Original RM} \label{subfig:RM1}
		\end{subfigure}
		\begin{subfigure}[b]{0.49\linewidth}
			\includegraphics[width=\linewidth]{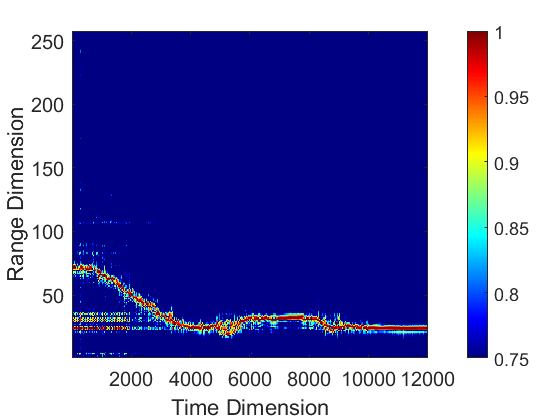}
			\caption{Normalized RM} \label{subfig:RM2}
		\end{subfigure}
		\begin{subfigure}[b]{0.49\linewidth}
			\includegraphics[width=\linewidth]{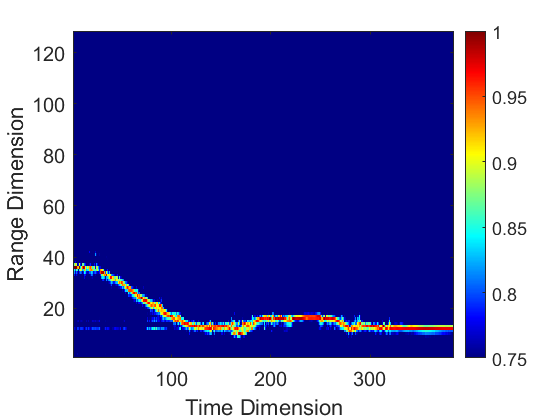}
			\caption{Resized RM} \label{subfig:RM3}
		\end{subfigure}
		\begin{subfigure}[b]{0.49\linewidth}
			\includegraphics[width=\linewidth]{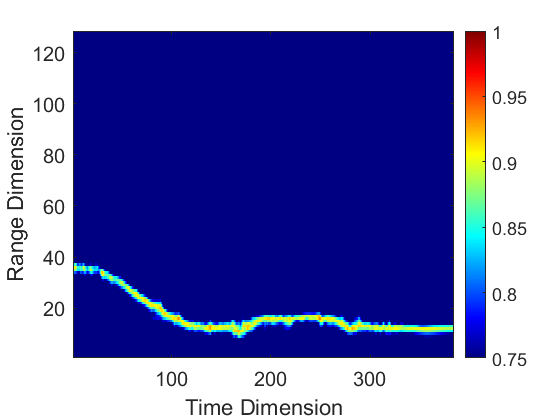}
			\caption{Filtered and thresholded RM} \label{subfig:RM4}
		\end{subfigure}
		\vspace{.5em}
		\caption{Range-map pre-processing steps.}
		\label{fig:RM_preprocessing}
	\end{figure}

	The range-map describes the target distance to the radar vs. slow-time. We consider the range-map as an image, $ x(m,n) $ with each sample converted to a decibel, absolute value. The radar data collected produces an image of size $ M \times N = 256 \times 12,000 $, where $ M $ and $ N $ represent the number of range bins (rows) and slow-times (columns), respectively.
	An initial step is to remove signal power dependency on range by column-wise normalization of the range-map. The strongest pixel $ x_{max}(n) $ of each column $ n $ is computed and used to divide all pixel values in that column. This step allows equal treatments of all ranges, despite their distances from the radar. Fig.~\ref{subfig:RM1} and \ref{subfig:RM2} show the range-map before and after normalization. In this case, the subject begins with translation motion, through walking, then follows with two consecutive in-place motions, namely,  sitting and standing. It is noted that, in this example, the signal-to-noise ratio is considerably high such that a simple thresholding can be used to "clean" the image and  highlight the target range signature. The normalized, resized, filtered, and thresholded range-map image is shown in Fig.~\ref{subfig:RM4}. Referring to the range-map, the following  equations describe the above  pre-processing steps. For normalization,

\begin{equation}
	\label{eq:normRM}
	R{M_{norm}}(n) = \frac{1}{{{x_{\max }}(m)}}RM(n)~,~n = 1,2,...,N
\end{equation}
\noindent It is noted that the original image size of $ 256 \times 12, 000  $ is excessively large, and thus can be reduced by down-sampling. We perform uniform sub-sampling over slow-time, and produce a smaller image size, referred to as $ RM_{ds} $,  of dimension $ M_{new} \times N_{new} = 128 \times 384 $. To improve the end result of the Radon transform, the down-sampled image $ RM_{ds} $ is filtered with a ($ 3\times3 $) smoothing kernel, given by,

\begin{equation}
	\label{eq:filterKernel}
	H = \frac{1}{9}\left[ {\begin{array}{*{20}{c}}
		1&1&1\\
		1&1&1\\
		1&1&1
		\end{array}} \right]
\end{equation}
		
\noindent The filtered image is expressed as,
	
\begin{equation}
	\label{eq:kernel2}
	R{M_{filtered}}(m,n) = \sum\limits_{i =  - k}^k {\sum\limits_{j =  - k}^k {H(i,j) \cdot R{M_{ds}}(m + i,n + j)} } ~,~ m =1,2,...,M_{new};~n =1,2,...,N_{new} 
\end{equation}
	
\noindent The final pre-processing step is applying a threshold. Values below a certain level in the above image equation are  set to zero, i.e., 

\begin{equation}
	\label{eq:RM_threshold}
	RM_{th}(m,n)= 
	\begin{cases}
	{R{M_{filtered}}(m,n)},~~\text{if } &{R{M_{filtered}}(m,n)}{ \ge 0.75}\\
	0,              &\text{otherwise}
	\end{cases}
\end{equation}
After performing normalization, resizing, filtering, and thresholding, we apply the Radon transform to the image in Fig.~\ref{subfig:RM4}.

\begin{figure}[hbtp]
		\centering
		\includegraphics[width=0.9\linewidth]{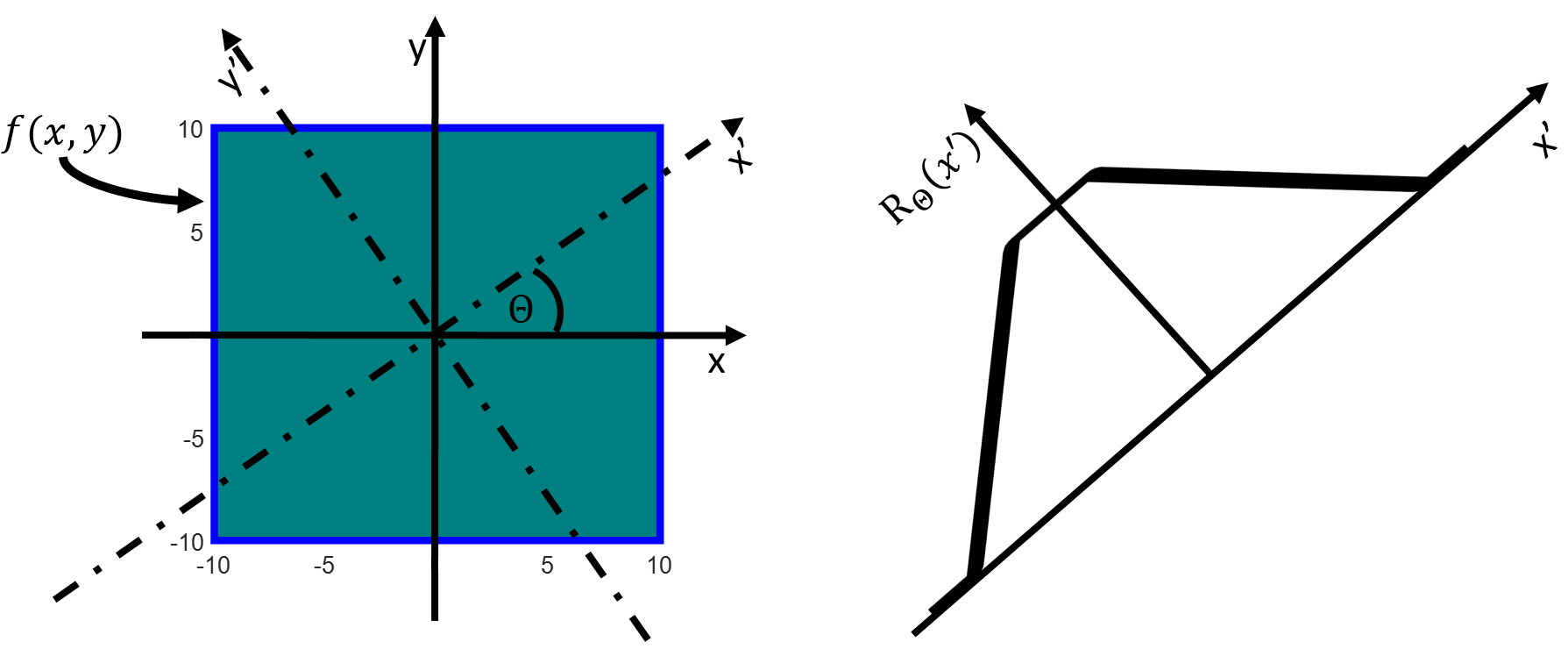}
		\vspace{.5em}
		\caption{Radon transform operations}
		\label{fig:radontransExampl}
	\end{figure}

	The Radon transform converts a Cartesian coordinate system $ (x,y) $ to an angle and a distance from a center point representation $ (\theta, x') $. It integrates all pixel values over  $ y' $. In other words, each pixel is projected onto the new subspace abscissa $ x' $. The projection point of every pixel on the $ x' $~axis can be rewritten as $ x' = x\cdot cos\theta + y\cdot sin\theta $. The Radon transform is graphically illustrated in Fig.~\ref{fig:radontransExampl}. The computing algorithm of Matlab divides each pixel into four subpixels. These subpixels are projected onto the subspace abscissa $ x' $ separately. Matlab separates the abscissa $ x' $ into bins. The subpixel which crosses a bin is counted with the full value. On the other side, if a subpixel hits the border of two bins, it will be split between these bins evenly \cite{26_helgason1980radon, 27_5751251, 28_mathworksRadonTrans}. The integration over the subspace ordinate axis $y'$ is given by, 
	
	\begin{equation}
	\label{eq:RadonEquation1}
	{R_\theta }(x') = \int\limits_{ - \infty }^\infty  {f(x,y)dy'} 
	\end{equation}
	
\noindent with,
		\begin{equation}
	\label{eq:RadonEquation2}
	\left[ {\begin{array}{*{20}{c}}
		{x'}\\
		{y'}
		\end{array}} \right] = \left[ {\begin{array}{*{20}{c}}
		{\cos \theta }&{\sin \theta }\\
		{ - \sin \theta }&{\cos \theta }
		\end{array}} \right] \cdot \left[ {\begin{array}{*{20}{c}}
		x\\
		y
		\end{array}} \right]
	\end{equation}
\noindent The inverse transform (rotation) of axes is given by,
		\begin{equation}
	\label{eq:RadonEquation3}
	\left[ {\begin{array}{*{20}{c}}
		{x}\\
		{y}
		\end{array}} \right] = \left[ {\begin{array}{*{20}{c}}
		{\cos \theta }&{- \sin \theta }\\
		{  \sin \theta }&{\cos \theta }
		\end{array}} \right] \cdot \left[ {\begin{array}{*{20}{c}}
		x'\\
		y'
		\end{array}} \right]
	\end{equation}
\noindent Accordingly, Eq.~\ref{eq:RadonEquation1} can be written as, 
	\begin{equation}
	\label{eq:RadonEquation4}
	{R_\theta }(x') = \int\limits_{ - \infty }^\infty  {f(x'\cos \theta  - y'\sin \theta ,x'\sin \theta  + y'\cos \theta )dy'} 
	\end{equation}

		\begin{figure}[hbtp] 
		\begin{subfigure}[b]{0.5\textwidth}
			\includegraphics[width=\textwidth]{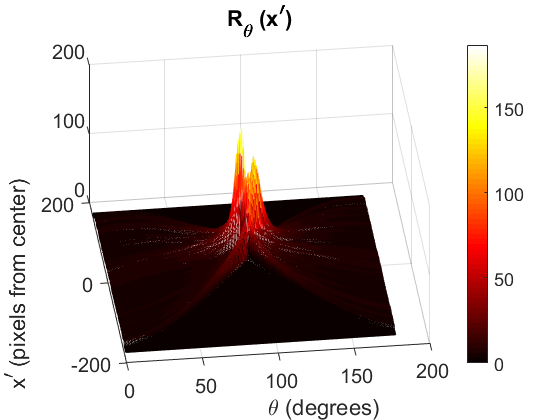}
			\caption{Radon transform - 3D representation }
			\label{subfig:RT1}
		\end{subfigure}
		\begin{subfigure}[b]{0.5\textwidth}
			\includegraphics[width=\textwidth]{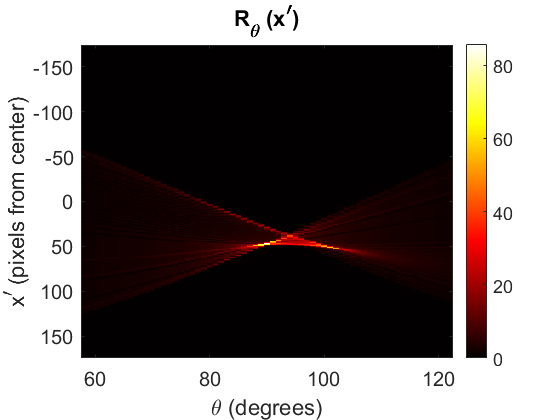}
			\caption{Radon transform - 2D representation}
			\label{subfig:RT2}
		\end{subfigure}
		\vspace{.5em}
		\caption{The Radon transform}  
		\label{fig:MeshGridRadonTrans}
	\end{figure}

\begin{figure}[hbtp] 
	\begin{subfigure}[b]{0.5\textwidth}
		\includegraphics[width=\textwidth]{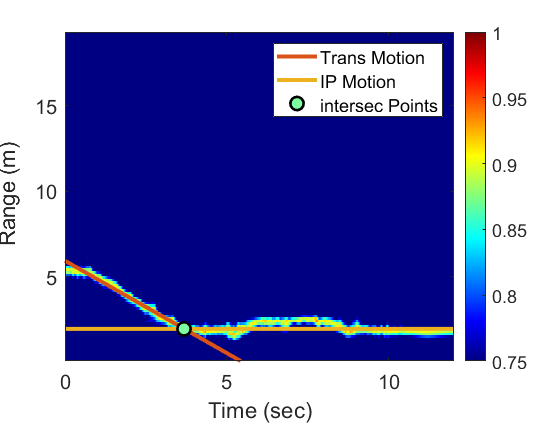}
		\caption{RM with intersection point}
		\label{subfig:dxProcess}
	\end{subfigure}
	\begin{subfigure}[b]{0.5\textwidth}
		\includegraphics[width=\textwidth]{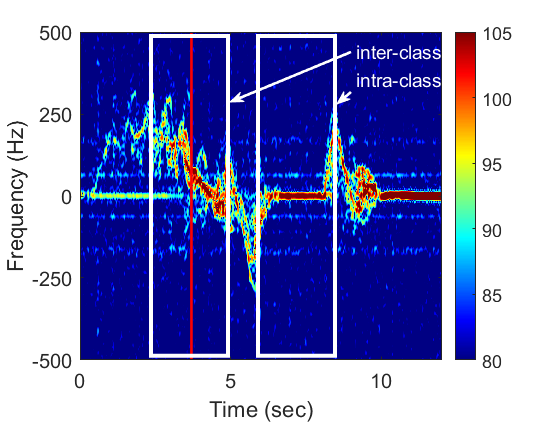}
		\caption{MD with intersection time (red line)}
		\label{subfig:MD_Separation}
	\end{subfigure}
			\vspace{.5em}
	\caption{ Walking towards the radar followed by sitting and then standing. } 
	\label{fig:MD_SeparationRonnySitting}
\end{figure}

We apply the Radon transform to the above example, specifically to the range-map of Fig.~\ref{subfig:RM4}. The best results were achieved by setting $\Delta\theta=1^\circ$. In examining the Radon transform in Fig.~\ref{fig:MeshGridRadonTrans}, it is clear that there is a peak at $\theta = 90^\circ$ angle, which refers to an in-place motion. The other peak off $\theta = 90^\circ$ refers to a translation motion. 
It is noted that the peak location defines both the angle $\theta$ and a distance from the origin $ x' $. According to this information, a linear function is computed by,
\begin{subequations}
	\begin{align} \label{eq:y_m_n_computing}
		y 		&= 	m(k)\cdot x+n(k)\\[.15in]
		\label{eq:m_computing}
		m(k) 	&= 	tan(\frac{\pi}{2} - \theta(k))=cot(\theta(k))\\[.15in]
		\label{eq:n_computing}
		n(k) 	&=  \frac{x'}{sin(\theta(k))}
	\end{align}
\end{subequations}
\noindent The index  $ k = 1,..,K-1 $ is the number of all detected motions. The slope $m(k)$ as well as the intersection point with the abscissa axis $n(k)$  of each linear function $y$ is computed by Eq.~{\color{blue}9b} and Eq.~{\color{blue}9c}, respectively.
With all linear functions in the range-map determined, the intersection point between the two consecutive lines, describing translation and in-place motions, can be simply obtained by solving the two linear equations, 
\begin{equation}
	\label{eq:intersecEquation}
\left[ {\begin{array}{*{20}{c}}
	{m(k)}&{ - 1}\\
	{m(k + 1)}&{ - 1}
	\end{array}} \right] \cdot \left[ {\begin{array}{*{20}{c}}
	x\\
	{y}
	\end{array}} \right] = \left[ {\begin{array}{*{20}{c}}
	{ - n(k)}\\
	{ - n(k + 1)}
	\end{array}} \right],k = 1,..,K - 1
\end{equation} 
\noindent Fig.~\ref{subfig:dxProcess} shows the intersection point of the two lines,  plotted over the $ RM_{th} $ of Fig.~\ref{subfig:RM4} as well as over the spectrogram, shown in Fig.\ref{subfig:MD_Separation}. The intersection time is very close to the true value when motion transitioned.

\subsection{Power Burst Curve (PBC)}
\label{sec:PBC}
	
From the previous section, the separation between in-place and translation motions can be  achieved by using the Radon transform applied to the range-map. To determine whether there is one or a sequence (multiple) of in-place motions, we need to examine the microDoppler signatures over the in-place motion interval. Referring to the spectrogram $S(n,k)$ in Fig.~\ref{subfig:MD_Separation}, the subject walks towards the radar in a translation motion, followed by sitting and standing activities, both of which represent in-place motions. In order to separate the two consecutive in-place motions, where range information is no longer a factor, we deal with the radar scattering as a deterministic signal, rather that a random process,\cite{31_536695} and measure the rise and fall of the signal energy in $S(n,k)$ over slow-time which is known as the power burst curve (PBC).\cite{32_7944316, 33_6889337}

\begin{figure}[hbtp]
	\centering
	\begin{subfigure}[b]{0.49\linewidth}
		\includegraphics[width=\linewidth]{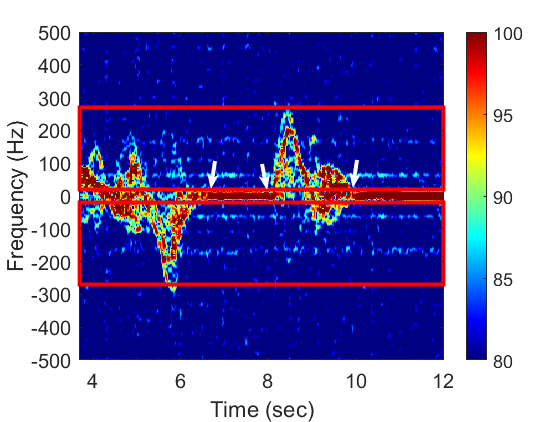}
		\caption{MD with Freq.-Bands} \label{subfig:PBC1}
	\end{subfigure} 
\begin{subfigure}[b]{0.49\linewidth}
		\includegraphics[width=\linewidth]{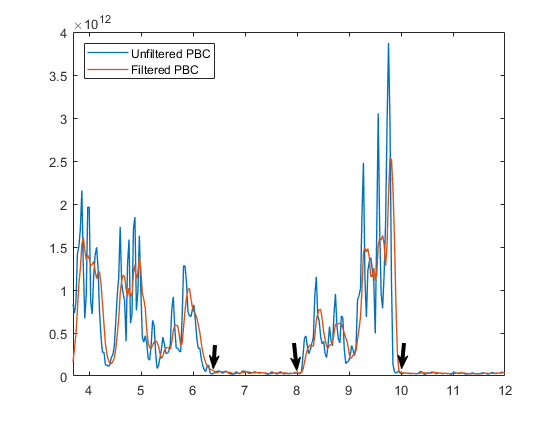}
		\caption{Unfiltered and filtered PBC} \label{subfig:PBC3}
	\end{subfigure} 
			\vspace{.5em}
	\caption{Illustration of PBC.}
	\label{fig:PBC}
\end{figure}

The Doppler frequencies at and around the zero Doppler axis are not of interest, since their strong power gain biases the PBC. In this work, the chosen frequency bands for power computation are between \(K_{P1} = 20Hz\) and \(K_{P2} = 270Hz\)  for  positive Doppler frequencies and  between 
\(K_{N2} =  - 20Hz\) and \(K_{N1} =  - 270Hz\) for negative Doppler frequencies. We compute the PBC for the combined frequency bands by, %
\begin{equation}
\label{eq:PBC}
S(n) = \sum\limits_{k_1 = {K_{P1}}}^{K_{P2}} {{{\left| {S(n,{k_1})} \right|}^2}}  + \sum\limits_{{k_2} = {K_{N1}}}^{{K_{N2}}} {{{\left| {S(n,{k_2})} \right|}^2}} ,n = 1,2,...,N
\end{equation}
In the above equation,  the index $n = 1,2,...,N$ represents the columns of the spectrogram. The PBC is drawn in Fig.~\ref{subfig:PBC3}.

The computations of Eq.~\ref{eq:PBC} results in a fluctuating power curve stemming from intricate microDoppler signatures of human motions. Such fluctuation may exhibit small values close to zero over a single motion time extent. These values could mistakenly define wrong event boundaries.

To mitigate the above problem, we apply a moving average filter.  The window size is set to $w=5$. The filtered PBC, $S_f(n,k)$ is computed by,
 
\begin{equation}
\label{eq:PBC_Filter}
{S_f}(n) = \frac{1}{w}\left( {{S}(n) + {S}(n - 1) + ... + {S}(w - n)} \right),n = 1,2,...,N
\end{equation}	
The filtered PBC is plotted in Fig.~\ref{subfig:PBC3}, and is used to determine the start and end of an activity. The threshold has been found empirically as $3\%$ over the minima ($S_{f_{min}}+ 0.03 \cdot(S_{f_{max}}-S_{f_{min}})$). Based on the this threshold value, the in-place activities are separated by,

\begin{equation}
\label{eq:PBC_threshold}
\bar{S}_f(n) = 
\begin{cases}
S_f(n),~~\text{if } & S_f(n) \ge threshold\\
end,              &\text{otherwise}
\end{cases}
\end{equation}

The application of PBC to the underlying example identified, with good approximation, the extent of each of the two in-place motions considered. This is shown by the corresponding arrows in the spectrogram and the PBC in Fig.~\ref{fig:PBC}. The combination of Radon transform and PBC is slated to avoid making classification decisions over time-windows which have inter-motion and intra-motion activities. The former refers to a window extending over two different classes, i.e., consecutive translation and in-place motions, whereas the latter represents the case of a window covering portions of two motions of the same class, i.e., consecutive in-place motions. Mixed motions within the same window confuses the classifier and may render decisions of neither motion at hand. The only way to remedy such confusion is to train the classifier with mixed activities which increases the complexity of the problem.  Examples of inter- and intra-motion windows are shown in Fig.~\ref{subfig:MD_Separation} by rectangulars.

\section{Conclusion}
\label{sec:conclusion}
In this paper, we presented a camera-less technique based on the Radon transform for detecting the transition time between translation and in-place motions, and vice versa. This technique uses the range-map that depicts range vs. slow-time of the moving target. The horizontal lines in the range-map correspond to in-place motions, whereas other lines describe a time-dependent range. The power burst curve was used over the interval of in-place motions to define the time boundaries to define the time boundaries of what could possibly be multiple motions, like sitting then standing. Finding the transition times as well as event time-span allows us to avoid rendering wrong motion classification decisions due to mixing two different motions within the same time window under testing. We provided an example that demonstrates the effectiveness of the proposed approach, which is deemed useful in dealing with continuous-time monitoring of human activities. Again, it is important to point out that from a Comcast Labs point of view, this is a technological test, not a product, and if any productization of this test would only occur, it would only be with the full, opt-in consent of the individual being monitored. 
	
\bibliography{report} 
\bibliographystyle{spiebib} 
\end{document}